\documentclass{ws-procs9x6}

\usepackage{graphicx}
\def\la{\mathrel{\mathchoice {\vcenter{\offinterlineskip\halign{\hfil
$\displaystyle##$\hfil\cr<\cr\sim\cr}}}
{\vcenter{\offinterlineskip\halign{\hfil$\textstyle##$\hfil\cr
<\cr\sim\cr}}}
{\vcenter{\offinterlineskip\halign{\hfil$\scriptstyle##$\hfil\cr
<\cr\sim\cr}}}
{\vcenter{\offinterlineskip\halign{\hfil$\scriptscriptstyle##$\hfil\cr
<\cr\sim\cr}}}}}
\def\ga{\mathrel{\mathchoice {\vcenter{\offinterlineskip\halign{\hfil
$\displaystyle##$\hfil\cr>\cr\sim\cr}}}
{\vcenter{\offinterlineskip\halign{\hfil$\textstyle##$\hfil\cr
>\cr\sim\cr}}}
{\vcenter{\offinterlineskip\halign{\hfil$\scriptstyle##$\hfil\cr
>\cr\sim\cr}}}
{\vcenter{\offinterlineskip\halign{\hfil$\scriptscriptstyle##$\hfil\cr
>\cr\sim\cr}}}}}

\begin{document}

\title{Evolution of optically faint AGN from COMBO-17 and GEMS}

\author{L. WISOTZKI, K. JAHNKE, S.F. SANCHEZ}

\address{Astrophysical Institute Potsdam, \\
An der Sternwarte 16, D-14482 Potsdam, Germany, lwisotzki@aip.de}

\author{C. WOLF}

\address{Department of Physics, University of Oxford, UK} 

\author{M. BARDEN, E.F. BELL, A. BORCH, B.~H\"AUSSLER, K. MEISENHEIMER, H.-W. RIX}

\address{Max-Planck-Institut f\"ur Astronomie, Heidelberg, Germany}

\author{S.V.W. BECKWITH, J.A.R. CALDWELL, S.~JOGEE, R.S. SOMERVILLE}

\address{Space Telescope Science Institute, Baltimore, USA}

\author{D.H. McIntosh}

\address{University of Massachusetts, Amherst/MA, USA}

\author{C.Y. Peng}

\address{Steward Observatory, University of Arizona, Tucson/AZ, USA}


\maketitle

\abstracts{We have mapped 
the AGN luminosity function and its evolution between $z=1$ and $z=5$ 
down to apparent magnitudes of $R<24$. Within the GEMS project
we have analysed HST-ACS images of many AGN in the 
Extended Chandra Deep Field South, enabling us to assess the evolution 
of AGN host galaxy properties with cosmic time.
}

\section{Introduction}

This article is a report on recent progress in the study of
optically faint Active Galactic Nuclei (AGN). Most of the
content has recently been published elsewhere; on the following
pages we provide a concise summary and present some of the
key figures.

\section{The AGN luminosity function from COMBO-17}

\begin{figure}[t]
\centerline{\includegraphics*[height=9cm,angle=-90]{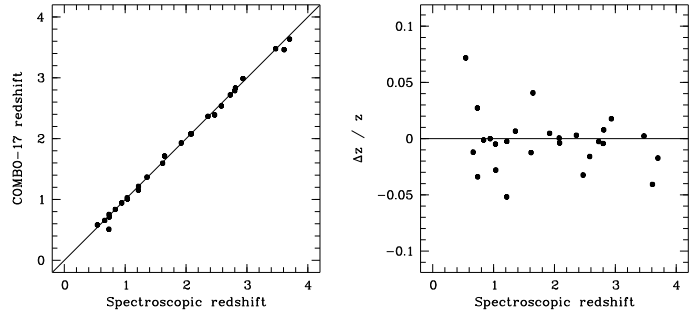}}
\caption{COMBO-17 spectrophotometric redshifts of AGN 
vs.\ spectroscopic redshifts.
\label{fig:dz}}
\end{figure}

The COMBO-17 survey (Wolf et al.\ 2004\cite{wolf*:04:COMBO17})
uses multi-band photometry in 17 filters within 
$350\,\mathrm{nm} \la \lambda_\mathrm{obs} \la 930\,\mathrm{nm}$.
By matching the photometry to an extensive template library,
we can simultaneously determine photometric redshifts of AGN with an 
accuracy of $\sigma_z<0.03$ (Fig.~\ref{fig:dz}), 
and obtain spectral energy distributions.
We have defined an AGN sample within $1.2 < z < 4.8$,
which implies that even at $z \simeq 3$, the sample reaches below 
luminosities corresponding to $M_B = -23$, conventionally employed 
to distinguish between Seyfert galaxies and quasars.

We clearly detect a broad plateau-like maximum of quasar activity around $z
\simeq 2$ and map out the smooth turnover between $z\simeq 1$ and $z\simeq 4$. 
The shape of the luminosity function is characterised by some mild curvature, but no
sharp `break' is present within the range of luminosities covered.
Using only the COMBO-17 data, the evolving LF can be adequately described 
by either a pure density evolution (PDE) or a pure luminosity evolution 
(PLE) model. However, the absence of a strong $L^*$-like feature in the shape 
of the LF inhibits a robust distinction between these modes.

\begin{figure}[ht]
\centerline{\includegraphics*[height=8cm,angle=-90]{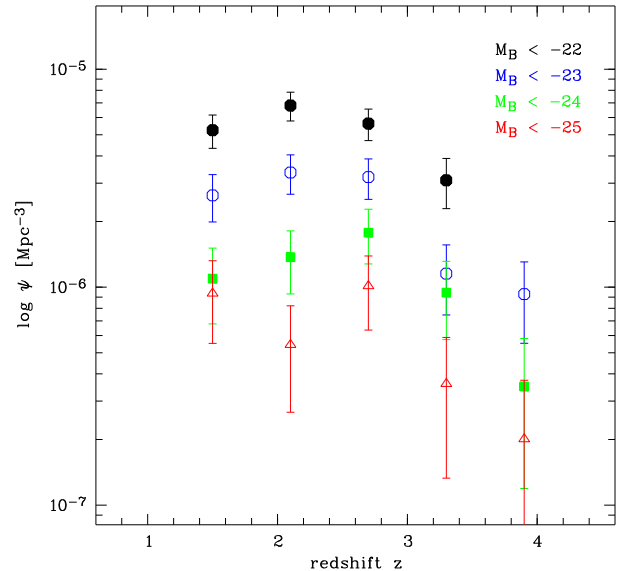}}
\caption{Space density of AGN in COMBO-17 as a function of redshift,
for different low-luminosity cutoffs.
\label{fig:sd}}
\end{figure}

We present a robust estimate for the integrated UV luminosity generation 
by AGN as a function of redshift. We find that the LF continues to rise 
even at the lowest luminosities probed by our survey, but that the slope 
is sufficiently shallow that the contribution of low-luminosity AGN 
to the UV luminosity density is negligible.
Although our sample reaches much fainter flux levels than previous 
data sets, our results on space densities and LF slopes are completely
consistent with extrapolations from recent major surveys such as SDSS and 2QZ.
Details of this analysis are published in the paper by 
Wolf et al.\ (2003\cite{wolf*:03:QLF}).

\section{Colors and stellar masses of AGN host galaxies at intermediate redshifts from GEMS}

GEMS is a two bands, F606W and F850LP, HST imaging survey of a continuous 
field of the `Extended Chandra Deep Field South', stretching over
$28' \times 28'$ in the sky (Rix et al.\ 2004\cite{rix04}). 
In this field, COMBO-17 provides SEDs
and redshifts of $\sim 10000$ galaxies and $\sim 100$ AGN.
We have constructed a subsample of these AGN by defining a redshift
slice, $0.5 < z < 1.1$, where the two GEMS bands bracket the rest-frame
4000~\AA\ break. We have detected the hosts of all these AGN 
in the F606W-band, recovering their fluxes, morphologies and structural 
parameters. A full account of this work is given by
Sanchez et al.\ (2004\cite{sanchez*:04:GHG1}).

\begin{figure}[ht]
\centerline{\includegraphics[angle=-90,width=8cm]{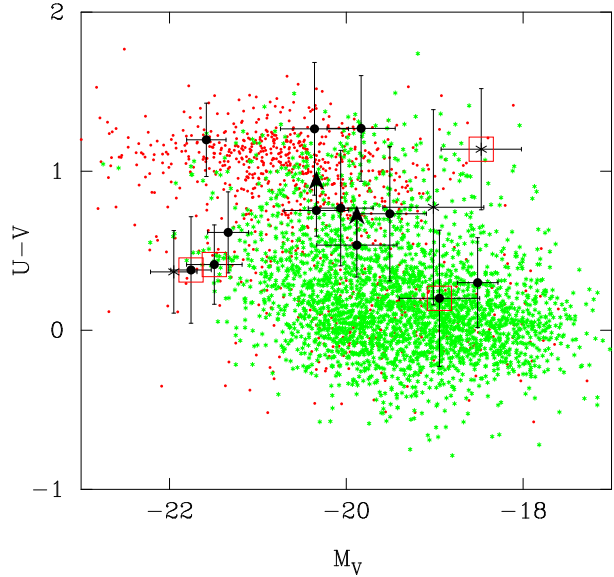}}
\caption[]{Colors and absolute magnitudes of AGN host galaxies from GEMS
within $0.5 < z < 1.1$. Filled symbols denote early-type systems,
open squares indicate interacting or merging objects.
For comparison, we show also the corresponding distribution of 
inactive galaxies at the same redshift range. 
The red sequence for elliptical galaxies is clearly identified with
$U-V \ga 0.8$.  The elliptical AGN hosts are clearly bluer, on average,
than the red sequence.
\label{fig:cmd}}
\end{figure}

\begin{figure}
\centerline{\includegraphics[angle=-90,width=8cm]{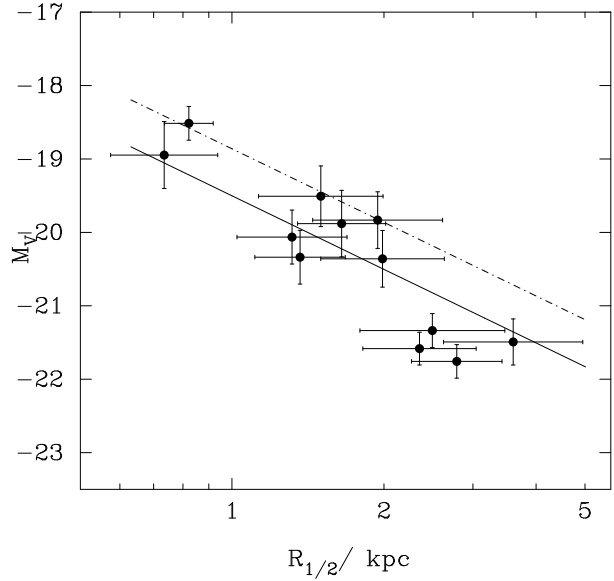}}
\caption[]{Host galaxy absolute magnitudes against half-light radii for the
  12 early-type galaxies at $0.5 < z < 1.1$ (filled circles). The solid line shows
  the luminosity-size relation for early-type red sequence
  galaxies at the mean redshift of our objects (Schade et al.\ 1997\cite{scha97}). 
  The dashed-dotted line shows the relation at $z=0$ from Kormendy (1977\cite{korm77}).
\label{fig:kormendy}}
\end{figure}

Morphologically, the AGN host galaxies are predominantly of early-type 
($\sim 80$~\%). Less than $\sim 20$~\% have structural properties
characteristic of late-type galaxies. The fraction of objects with
disturbed morphological appearance suggestive of ongoing galaxy 
interactions is also $\sim 20$~\%. The hosts
show a wide range of colors, from being as red as red sequence galaxies
to colors as blue as galaxies undergoing star formation. 
Comparing with single stellar population models,
the average stellar population would have an age of $\sim 1$~Gyr.  
With $\sim 70$~\% of the objects having $U-V<0.8$, the early-type AGN 
hosts are significantly bluer than red sequence early-type galaxies
(see Fig.~\ref{fig:cmd}).
However, their color-magnitude distribution is consistent with the 
distribution of \emph{all} inactive early-type galaxies in the GEMS field 
when also the blue tail of these objects is taken into account.

Despite their sometimes very blue colors, the early-type AGN hosts
are structurally similar to red sequence ellipticals: They
follow the Kormendy relation (Fig.~\ref{fig:kormendy}), and 
their absolute magnitudes ($M_V \sim -20.2$), effective radii 
($r_{1/2} \sim 2$~kpc) and stellar masses 
($\sim 10^{10}$--$10^{11}\:M_\odot$) are in the range of normal
ellipticals.

\section{UV light in QSO host galaxies at $1.8 < z < 2.75$}

We have exploited GEMS to investigate a sample of 23 AGN in
the redshift range $1.8<z<2.75$, also drawn from the COMBO-17 
survey. In 9 of the 23 AGN we resolve the host galaxies in both 
filter bands, whereas in the remaining 14 objects,
any resolved components have less than 5~\% of the nuclear flux 
and were considered nondetections. However, when we coadd the
unresolved AGN images into a single high signal-to-noise composite
image we find again an unambiguously resolved host galaxy.
The recovered host galaxies have apparent magnitudes of
$23.0<\mathrm{F606W}<26.0$ and $22.5<\mathrm{F850LP}<24.5$ 
with rest-frame UV colours in the range
$-0.2<(\mathrm{F606W}-\mathrm{F850LP})_\mathrm{obs}<2.3$. 
The rest-frame absolute magnitudes at 200~nm are
$-20.0<M_{200~\mathrm{nm}}<-22.2$. The photometric properties of the
composite host are consistent with the individual resolved host
galaxies.

\begin{figure}[ht]
\centerline{\includegraphics[clip,angle=-90,width=8cm]{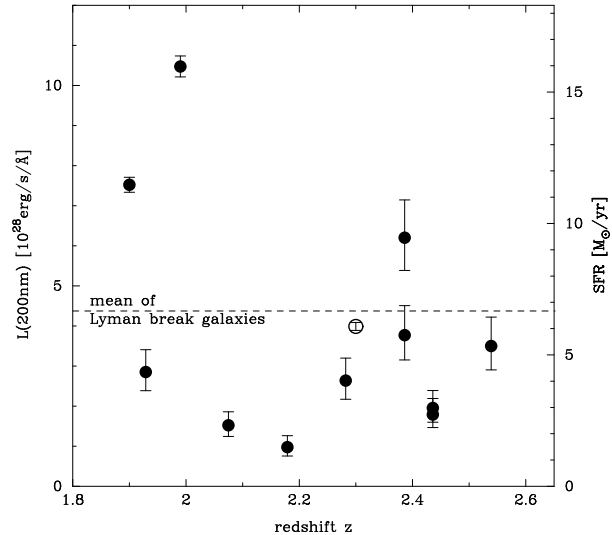}}
\caption[]{Rest frame 200nm luminosities, and star formation rates as derived
from from the F606W-band, both uncorrected for dust. The open symbol
marks the SFR of the `stacked' object created from the AGNs with
individually unresolved host galaxies. The horizontal dashed line is
the value obtained by Erb et al.\ (2003\cite{erb03}) for Lyman break galaxies at
$z=2.5$.
\label{fig:hz}}
\end{figure}

We find that the UV colors of all host galaxies are
substantially bluer than expected from an old population of stars with
formation redshift $z\le5$, independent of the assumed
metallicities. These UV colours and luminosities range up to the
values found for Lyman-break galaxies (LBGs) at $z=3$. The presence
of significant amounts of UV light 
suggest either a recent starburst, of e.g.\ a few per cent of the
total stellar mass and 100~Myrs before observation, with mass-fraction
and age strongly degenerate, or ongoing star formation.
For the latter case we estimate star formation rates of typically
$\sim$$6\,\mathrm{M}_\odot\;\mathrm{yr}^{-1}$ (uncorrected for
internal dust attenuation), which again lies in the range of rates
implied from the UV flux of LBGs. For details see our recently
submitted paper (Jahnke et al.\ 2004\cite{jahnke*:04:GHG2}).

\end{document}